\documentclass[aps,preprintnumbers,eqsecnum,amsmath,amssymb,nofootinbib]{revtex4}
\usepackage{graphics}
\usepackage{dcolumn}
\usepackage{bm}
\usepackage{epsfig}
\usepackage{float}

\begin{document}

\title{\boldmath Constraints on the anomalous $Wtb$ couplings from $B$-physics experiments}
\author{Anastasia Kozachuk$^{a}$ and Dmitri Melikhov$^{a,b,c}$}
\affiliation{
$^a$D.~V.~Skobeltsyn Institute of Nuclear Physics, M.~V.~Lomonosov Moscow State University, 119991, Moscow, Russia\\
$^b$Joint Institute for Nuclear Research, 141980 Dubna, Russia\\
$^c$Faculty of Physics, University of Vienna, Boltzmanngasse 5, A-1090 Vienna, Austria}
\begin{abstract}
We analyze constraints on the anomalous $Wtb$ couplings from $B$-physics experiments, performing a correlated analysis and 
allowing all anomalous couplings to differ simultaneously from their Standard-Model (SM) values. 
The $B$-physics observables allow one to probe three linear combinations out of the four anomalous couplings, 
which parameterize the $Wtb$ vertex under the assumption that the SM symmetries remain the symmetries of the effective theory. 
The constraints in this work are obtained by taking into account the following $B$-physics observables: 
the $\bar B^0-B^0$ oscillations, the leptonic $B\to \mu^+\mu^-$ decays, the inclusive radiative $B\to X_s\gamma$ decays, 
and the differential branching fractions in the semileptonic inclusive $B\to X_s \mu^+ \mu^-$ and 
exclusive $B\to (K,K^*)\mu^+\mu^-$ decays at small $q^2$, $q$ the momentum of the $\mu^+\mu^-$ pair. 
We find that the SM values of the anomalous couplings belong to the 95\% CL allowed region obtained this way but lie 
beyond the 68\% allowed region. We also report that the distributions of the anomalous couplings obtained within our 
scenario differ from the results of the 1D scenario, when only one of the couplings is allowed to deviate from its SM value.
\end{abstract}
\maketitle
 
\section{Introduction}
The top quark is the heaviest of all known elementary particles and may be expected to have 
large couplings with physics beyond the Standard Model (BSM). One of the possibilities 
to probe BSM physics is to study the anomalous structure of the $Wtb$ vertex. 

At the scale $\mu\sim M_W$, the effects of BSM interactions may be parameterized in the framework of an effective field theory 
by a tower of higher-dimension operators constructed from the SM fields and obeying the SM symmetries \cite{Buchmuller}.  
The most general Lagrangian of the $Wtb$ vertex has the following form \cite{Kane}   
\begin{eqnarray}
{\cal L}_{Wtb}&=&{\cal L}_{Wtb}^{\rm SM}+\frac{g}{\sqrt{2}}  
\left[
\bar{b}\gamma^{\mu}\left(f_{VL}P_L + f_{VR}P_R\right) t\,W_{\mu} -
\bar{b}\frac{\sigma^{\mu\nu} \partial_{\nu} W_{\mu}}{M_W}\left(f_{TL}P_L + f_{TR}P_R\right)t \right] + \rm{h.c.},\nonumber \\
&&{\cal L}_{Wtb}^{\rm SM}=V_{tb}^*\frac{g}{\sqrt{2}}\bar{b}\gamma^{\mu}P_L t\,W_{\mu} + \rm{h.c.}, 
\label{L_Wtb}
\end{eqnarray}
where $P_{L/R}=\frac{1}{2}(1\mp\gamma_{5})$, $\sigma_{\mu\nu}=\frac{i}{2}(\gamma_{\mu}\gamma_{\nu} -\gamma_{\nu}\gamma_{\mu})$,  
$g$ is the $SU(2)$ gauge coupling. The Lagrangian (\ref{L_Wtb}) corresponds to the choice 
of the covariant derivative acting on a left quark doublet with weak hypercharge $Y$ in the form 
\begin{eqnarray}
D_\mu=\partial_\mu-i\frac{g}{2}\tau^a W_\mu^a-i\frac{g'}{2}Y B_\mu. 
\end{eqnarray}
The anomalous couplings $f_{VL}$, $f_{VR}$, $f_{TL}$ and $f_{TR}$ are induced by dimension-6 operators of the effective theory. 
In the SM, all anomalous couplings have zero values: $f_{VL} = f_{VR} = f_{TL} = f_{TR} = 0$.\footnote{For convenience, 
we provide a comparison with the anomalous couplings used in other papers: 
One finds $f_{VL}=V_L$ \cite{fajfer1}$=v_L-1$ \cite{misiak}$=f_{1}^L-1$ \cite{Kane}. For other couplings, 
$f_{VR}=V_R$ \cite{fajfer1}$=v_R$ \cite{misiak}$=f_{1}^R$ \cite{Kane}, 
$f_{TL}=G_L$ \cite{fajfer1}$=g_L$ \cite{misiak}$=f_{2}^L$ \cite{Kane}, and 
$f_{TR}=G_R$ \cite{fajfer1}$=g_R$ \cite{misiak}$=f_{2}^R$ \cite{Kane}.} 
Notice that vanishing of the anomalous couplings in the SM Lagrangian 
does not forbid the appearance of the corresponding Lorentz structures in the $t\to W b$ amplitude through radiative corrections \cite{koerner1}. 

The $SU_L(2)\otimes U_Y(1)$ gauge invariance leads to the appearance of the anomalous structures not only in the $Wtb$, but also
in other $Wqq'$ vertices \cite{misiak}. Imposing a number of constraints (e.g., no tree-level FCNC), 
these structures are given in terms of the same anomalous couplings as in Eq.~(\ref{L_Wtb}) and the appropriate 
CKM matrix elements. 

The anomalous couplings are in general complex numbers \cite{fajfer1,koerner1,koerner2,Wtb_DeVries}; this paper studies the constraints on the couplings in a restricted scenario 
assuming that they are real quantities. 
 
Obtaining experimental bounds on the anomalous couplings is an important direction in the search for New Physics. 
Such bounds may be obtained from different sources: e.g., from the direct production of top-quarks at hadron colliders, where 
weak processes is the main mechanism of the single $t$-production \cite{cms}. 
Another promising way is the indirect probe of the anomalous couplings from FCNC processes in $B$-physics: here, virtual top 
often gives the leading contribution, thus opening 
the possibility to constrain its anomalous couplings 
\cite{misiak, fajfer1, fajfer2}. This paper follows the line of the analysis of \cite{fajfer1} and studies more closely the correlations 
between the anomalous couplings that can be obtained from the $B$-physics data. 


\section{Effective Lagrangian for weak FCNC $B$-decays}
For the description of weak $B$-decays, an appropriate physics 
scale is $\mu\simeq 5$ GeV; all particles with much heavier masses
are not dynamical and may be integrated out within the formalism based on the operator product expansion.
The light degrees of freedom, such as the $u,d,c,s,b$ quarks, the photon and the gluons are dynamical degrees of freedom. 
For FCNC $B$-decays, this approach leads to the 
effective Lagrangian which involves operators built up of the light degrees of freedom 
(for details and the full set of the basis operators see \cite{Grinstein,Buras1,Buras2} and the discussion in 
\cite{altmannshofer,bobeth1,bobeth2}): 
\begin{eqnarray}
\label{Heffb2s}
{\cal L_{\rm eff}}(b\to s)=
\frac{G_F}{\sqrt{2}}V_{tb}V_{ts}^*\sum_{i=1}^{10}C_i{\cal O}_i.
\label{Heff} 
\end{eqnarray}
The operators ${\cal O}_{1-6}$ in (\ref{Heff}) are four-quark operators containing quark fields $\bar s$, $b$,  $\bar c$, $c$ 
with different color contractions. One also has the analogous four-quark operators with $c$-quark replaced by $u$-quark; 
the corresponding Wilson coefficients are however strongly CKM suppressed as $V_{ub}V^*_{us} \ll V_{tb}V^*_{ts}\simeq - V_{cb}V^*_{cs}$. 
For instance, the operators generating the charming loops are 
\begin{eqnarray}
{\cal O}_1=\bar s^j \gamma_\mu(1-\gamma_5)c^i \,\bar c^i \gamma^\mu(1-\gamma_5)b^j, \qquad 
{\cal O}_2=\bar s^i \gamma_\mu(1-\gamma_5)c^i \,\bar c^j \gamma^\mu(1-\gamma_5)b^j.   
\end{eqnarray}
The operators ${\cal O}_{7-10}$ are bilinears in quark fields 
\begin{eqnarray}
\label{operators}
{\cal O}_7=\frac{e m_b}{8\pi^2}(\bar s\sigma_{\mu\nu}(1+\gamma_5)b)F^{\mu\nu}, \quad 
{\cal O}_8=\frac{g_s m_b}{8\pi^2}(\bar s\sigma_{\mu\nu}(1+\gamma_5)T^A b)G_A^{\mu\nu},\nonumber \\
{\cal O}_9=\frac{e^2}{8\pi^2}(s\gamma_\mu(1-\gamma_5)b)(\bar l \gamma_\mu l),\quad 
{\cal O}_{10}=\frac{e^2}{16\pi^2}(s\gamma_\mu(1-\gamma_5)b)(\bar l \gamma_\mu\gamma_5 l). 
\end{eqnarray}
The $\mu$-dependent Wilson coefficients $C_i(\mu)$ at the scale $\mu\sim M_W$ encode the effects of heavy particles; 
in the SM these are $M_W$, $Z$, and $t$-quark, but in the extensions of the SM other heavy particles contribute thus changing the 
values of the Wilson coefficients. One can write \begin{eqnarray}
C_i=C_i^{SM}+\delta C_i, 
\end{eqnarray}
where the additional terms $\delta C_i$ reflect the New Physics contributions. The SM Wilson coefficients have the values: $C^{SM}_2(M_W)=1$, 
$C^{SM}_9(\mu\simeq m_b)=4.21$, $C^{SM}_{10}(\mu\simeq m_b)=-4.41$, $C^{SM}_7(\mu\simeq m_b)=-0.32$. 
We do not consider in our analysis the operators ${\cal O}^{'}_{7,9,10}$ with the opposite chirality compared to ${\cal O}_{7,9,10}$, respectively: 
In the SM, $C^{'}_7=C_7 m_s/m_b$, $C^{'}_{9,10}=0$, and the corrections $\delta {C}^{'}_{7,9,10}$ induced 
by the anomalous couplings are proportional to $m_s/m_b$ and are thus much smaller than $\delta {C}_{7,9,10}$.

The additions to the Wilson coefficients $\delta {C}_{7,9,10}$
due to the anomalous couplings were obtained in \cite{misiak,fajfer1}; they are linear functions of the following anomalous couplings:
\begin{eqnarray}
\delta C_7(f_{VL},f_{VR},f_{TL},f_{TR}), \qquad 
\delta C_9(f_{VL},f_{TR}), \qquad \delta C_{10}(f_{VL},f_{TR}). 
\end{eqnarray}
For the explicit formulas, we refer to Appendix B of Ref.~\cite{fajfer1} and we apply those formulas for the matching scale $\mu=2M_W$. 
We also use the $B$-$\bar B$ oscillations; the correction to the corresponding amplitude due to the anomalous couplings was calculated 
in \cite{fajfer2}. Noteworthy, it contains only two anomalous couplings, $f_{VL}$ and $f_{TR}$. 

$B$-meson observables, in addition to the Wilson coefficients, involve the amplitudes of the effective operators; 
the latter contain complicated hadron effects. So, in practice, only those processes involving
$B$-mesons, where the hadron uncertainties are kept under reasonable theoretical control, may be used for probing the
anomalous $Wtb$ couplings. A plausible strategy for the purpose of searching New Physics (NP) is to avoid, e.g., 
hadronic weak $B$-meson decays. 

An unavoidable difficulty in the theoretical description of FCNC $B$-decays is the calculation of the contributions 
generated by the four-quark operators 
in (\ref{Heff}), the so-called charming loops, especially their nonfactorizable parts \cite{charm1,charm2,charm3,charm4}. 
Recall, that the contribution of the virtual charm to FCNC $B$-decays is not CKM suppressed compared to the contribution of the top-quark. 
In the amplitudes of $B\to (K,K^*)l^+l^-$ decays, the factorizable charm contribution 
is governed by the linear combination of the Wilson coefficients $C_2+3C_1$ which is strongly scale-dependent. 
For instance, at $\mu\simeq m_b$,
$C_2+3C_1=0.3$ to be compared with the value $C_{9V}=4.21$ that governs a part of the top-quark contribution. Indeed, at small $q^2$, $q$ 
the momentum of the $l^+l^-$ pair, there in a sizeable numerical suppression of the charming loops compared to the top contribution, 
but this suppression is the subject to the precise choice of the scale $\mu$, indicating the importance of higher-order QCD corrections. 
As $q^2$ increases, the charm contribution rises much faster than that of the top; the charm takes over the top 
when one approaches the charmonia region \cite{charm1}. Finally, for the purpose of the search for NP, one can use FCNC $B$-decays below the 
charm threshold as soon as one has reliable theoretical predictions for the $B\to (K,K^*)$ form factors and for the contribution of 
the charming loops. 

For our further analysis, it is important to pay attention to the following facts: 
\begin{itemize}
\item[(i)] the coefficients $\delta C_{9}$ and $\delta C_{10}$, as well as the amplitude of the $B$-$\bar B$ oscillations 
do not get contributions from $f_{VR}$ and $f_{TL}$. 
\item[(ii)] the coefficient $\delta C_7$ involves only one linear combinations of $f_{VR}$ and $f_{TL}$.
\end{itemize}
So, in practice, precision measurements of $\delta C_{i}$ ($i=7,9,10$) from $B$-physics allows one to get access to 
$f_{VL}$, $f_{TR}$ and one specific linear combination of $f_{VL}$ and $f_{TR}$, determined by $\delta C_7$. 


\section{Bounds on anomalous $Wtb$ couplings}

When studying constraints on the anomalous couplings, both from direct top-quark production and from $B$-physics, 
one often considers different scenarios, depending on how many couplings are allowed to vary from their zero SM values. 
For instance, one-dimensional scenarios, when only one of the couplings is allowed to be nonzero, are well known \cite{misiak,fajfer1,cms}.  
Using such an approach, however, one does not access those regions where different anomalous couplings may have strongly correlated 
values far away from their zero SM values. We therefore follow here a different strategy: we allow all couplings to be nonzero and obtain the corresponding bounds from the data. 

As already noticed above, only those $B$-decay channels, where theoretical QCD uncertainties are kept under control, 
may be used for the extraction of the anomalous couplings. We use for our analysis the following channels:
\begin{itemize}
\item $B_{s}-\bar{B}_{s}$ oscillations (see e.g. \cite{lenz}): We consider the ratio of $\Delta M_{s}=M_{B_s}-M_{\bar B_s}$ in the theory with anomalous couplings over $\Delta M_{s}$ in the SM; its dependence on the anomalous couplings was calculated in \cite{fajfer2}. We fit this quantity to the ratio of the experimental $\Delta{M_s}$ \cite{HFLAV} over its lattice determination \cite{FLAG}. 
\item ${\rm Br}(\bar{B}\to X_s\gamma)|_{E_\gamma>1.6\,{\rm GeV}}$: 
data from \cite{HFLAV} and theoretical estimates from \cite{BXsgamma_theory}. 
\item ${\rm Br}(\bar{B}\to X_s\mu^+\mu^-)_{{\rm low}-q^2}$, $q$ the momentum of the $\mu^+\mu^-$ pair: 
data from \cite{HFLAV} and theoretical inputs from \cite{fajfer1}.
\item
$B_s\to \mu^+\mu^-$: 
data from \cite{HFLAV,Bmumuexp} and the theoretical predictions from 
\cite{BXsgamma_theory} (see also \cite{Bmumuth}). 

\item    
${\rm Br}(\bar{B}\to K^*\mu^+\mu^-)_{{\rm low}-q^2}$:  
data from \cite{BKstarmumuexp} in the lowest $q^2$-bins: $q^2=[0.1,1]$ GeV$^2$ and  $q^2=[1.1,6.0]$ GeV$^2$.
The differential distributions and asymmetries for these decays have been calculated in \cite{Altmannshofer2009}; 
convenient formulas parameterizing the theoretical results as functions of the Wilson coefficients were given in \cite{BKstarmumuth}. 
In this channel, the accuracy of the experimental results is slightly better than that of the theoretical predictions: 
the $B\to K^*$ form factors at small $q^2$ necessary for calculating the differential branching fractions of interest come 
mainly from light-cone QCD sum rules (LCSR); this method unfortunately does not allow a solid control over the systematic uncertainties 
of the calculated form factors \cite{m1,m2} (cf. \cite{lcsr1,lcsr2}). 
We assign a 15\% uncertainty on the LCSR predictions for the $B\to (K,K^*)$ form factors, 
yielding a 30\% uncertainty on the differential distributions. 

\item   
${\rm Br}(\bar{B}^+\to K^+\mu^+\mu^-)_{{\rm low}-q^2}$: we make use of the LHCb data \cite{LHCbB2K1} in two bins, $q^2=[0.1,0.98]$ GeV$^2$ and 
$q^2=[1.1,6.0]$ GeV$^2$ (see also \cite{LHCbB2K2}). 
The necessary $B\to K$ form factors come from LCSRs, e.g. \cite{lcsr1,lcsr2,charm1}. 
The most recent analysis \cite{lcsr2} yields the form factors 
$f_+^{B\to K}(0)=0.27\pm 0.08$ and $f_T^{B\to K}(0)=0.25\pm 0.07$. These numbers are somewhat smaller than 
those from the previous analysis \cite{charm1}:  
$f_+^{B\to K}(0)=0.34^{+0.05}_{-0.02}$ and $f_T^{B\to K}(0)=0.39^{+0.34}_{-0.03}$. 
Noteworthy, also the uncertainties reported in \cite{lcsr2} 
are considerably larger compared to the previous estimates. 
Taking into account the already mentioned problem with the assignment of uncertainties within QCD sum rules, 
for the analysis we employ the $B\to K$ form-factor parametrizations from \cite{latticeB2K}, 
which correspond to $f_+^{B\to K}(0)=0.33$ and $f^{B\to K}_T(0)=0.28$, and assign to them a 15\% theoretical uncertainty. 
The full contribution of the charming loops, including the factorizable and the nonfactorizable effects, from \cite{charm1} is taken 
into account. Convenient expressions for $10^8\,d{\rm Br}(B\to K\mu^+\mu^-)/dq^2$ [in GeV$^{-2}$] as functions 
of the additions to the Wilson coefficients in two bins of low-$q^2$ are given below: 
\begin{eqnarray}
q^2=[0.1,1.0]\,{\rm GeV}^2: &&\nonumber \;\\
10^8\,\frac{d{\rm Br}(B\to K\mu^+\mu^-)}{dq^2} [GeV^{-2}]&=&
41.14-12.01 \delta C_{10}+1.36  \delta C_{10}^2+1.42 \delta C_7^2\\
&&+7.33 \delta C_9+0.91 \delta C_9^2 +\delta C_7 (9.15+2.28\delta C_9),\nonumber\\
q^2=[1.1,6.0]\,{\rm GeV}^2: &&\nonumber\; \\
10^8\,\frac{d{\rm Br}(B\to K\mu^+\mu^-)}{dq^2}  [GeV^{-2}]&=&
34.33
-8.61 \delta C_{10}+0.98  \delta C_{10}^2
+1.55 \delta C_7^2\nonumber\\
&&+ 7.46 \delta C_9+0.92 \delta C_9^2 +\delta C_7 (9.73+2.37\delta C_9).
\end{eqnarray}
\end{itemize}
The amplitudes of these processes involve, via $\delta C_7$,  $\delta C_9$ and $\delta C_{10}$,  
the anomalous couplings $f_{VL}$, $f_{TR}$, but only one linear combination of the couplings $f_{VR}$ and $f_{TL}$: 
\begin{eqnarray}
f\equiv a_{\rm VR}f_{\rm VR}+ a_{TL}f_{TL},
\end{eqnarray}
with $a_{VR}\simeq 98.6$ and $a_{TL}\simeq -50.1$ at the scale 5 GeV \cite{fajfer1}. 
In what follows we obtain experimental bounds on $f$, $f_{VL}$, and $f_{TR}$ at the scale $\mu=2M_W$ from the $B$-physics data listed above. 

With the theoretical expressions for the observables of interest as functions of the anomalous couplings at hand, we proceed as follows: 
for a combination of the independent observables $A_j$ with the calculated theoretical dependence on the 
set of the anomalous couplings $F$ in the form $A_{j}(F)$, the experimental averages $\overline{A}_j$,
and the uncertainties $\Delta A_j$, we obtain the combined probability distribution of $F$ as follows
\begin{eqnarray}
\label{3D}
\rho(F)\propto \prod\limits_{j}
\exp
\left[
-\frac12\left(
\frac{A_j(F)-\overline{A}_j}{\Delta A_j}
\right)^2
\right].
\end{eqnarray}
where $\Delta A_j$ is obtained by combining the experimental and the theoretical uncertainties 
of the corresponding observable as $(\Delta A_j)^2\equiv (\Delta A_{j,th})^2+(\Delta A_{j,exp})^2$. 

\begin{figure}[h!]
\begin{center}
\includegraphics[width=7cm]{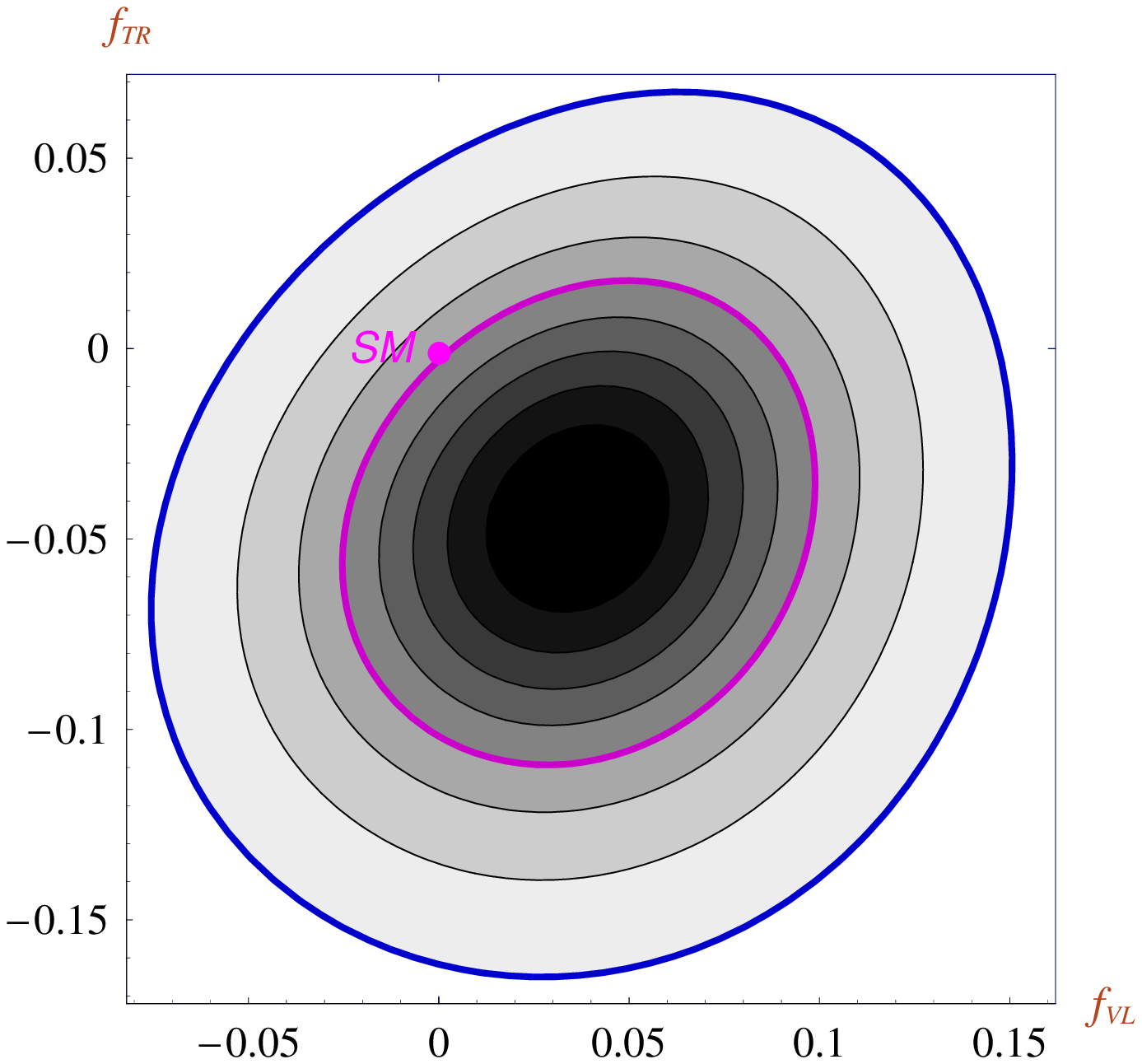}
\includegraphics[width=7cm]{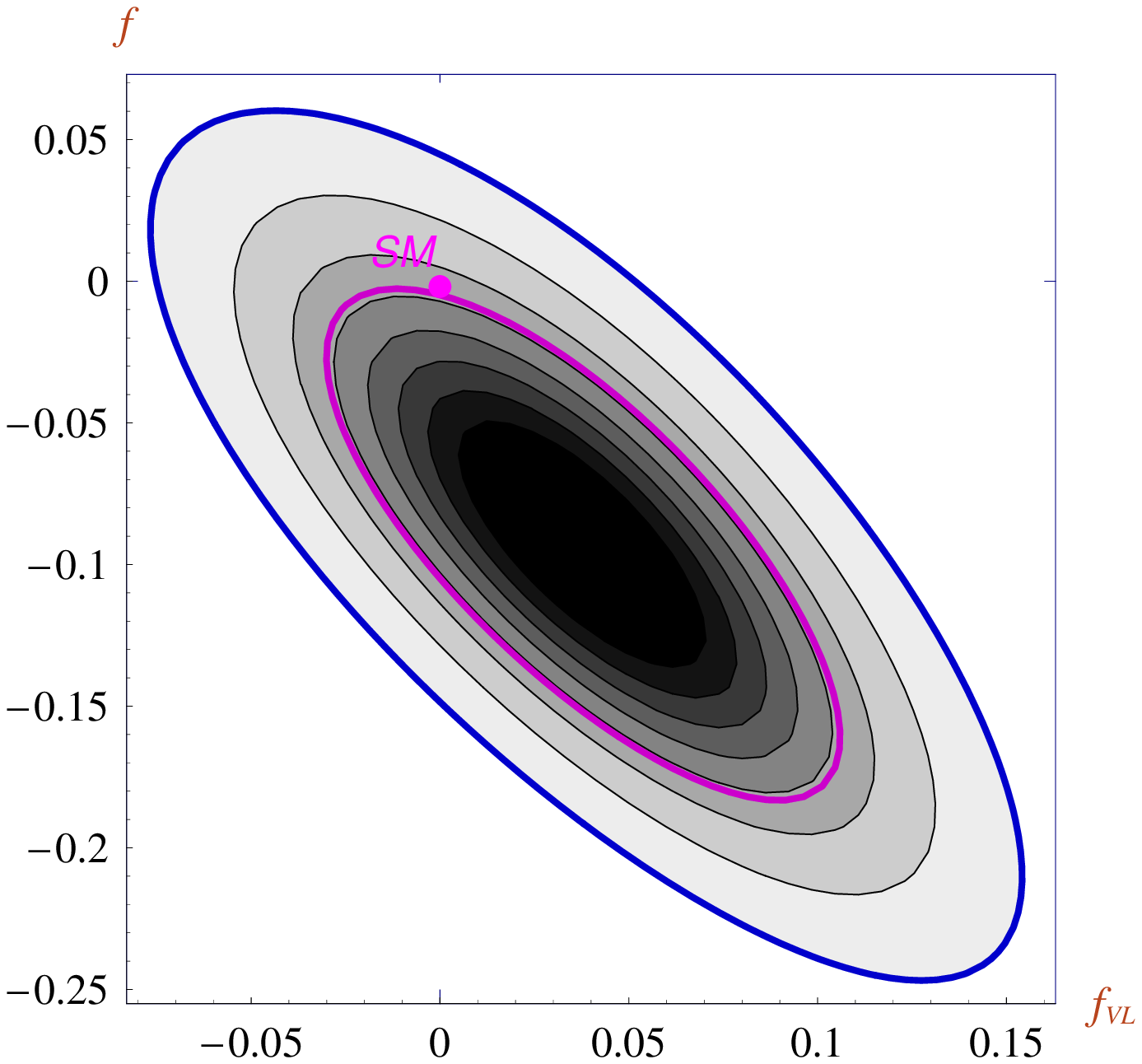} 
\includegraphics[width=7cm]{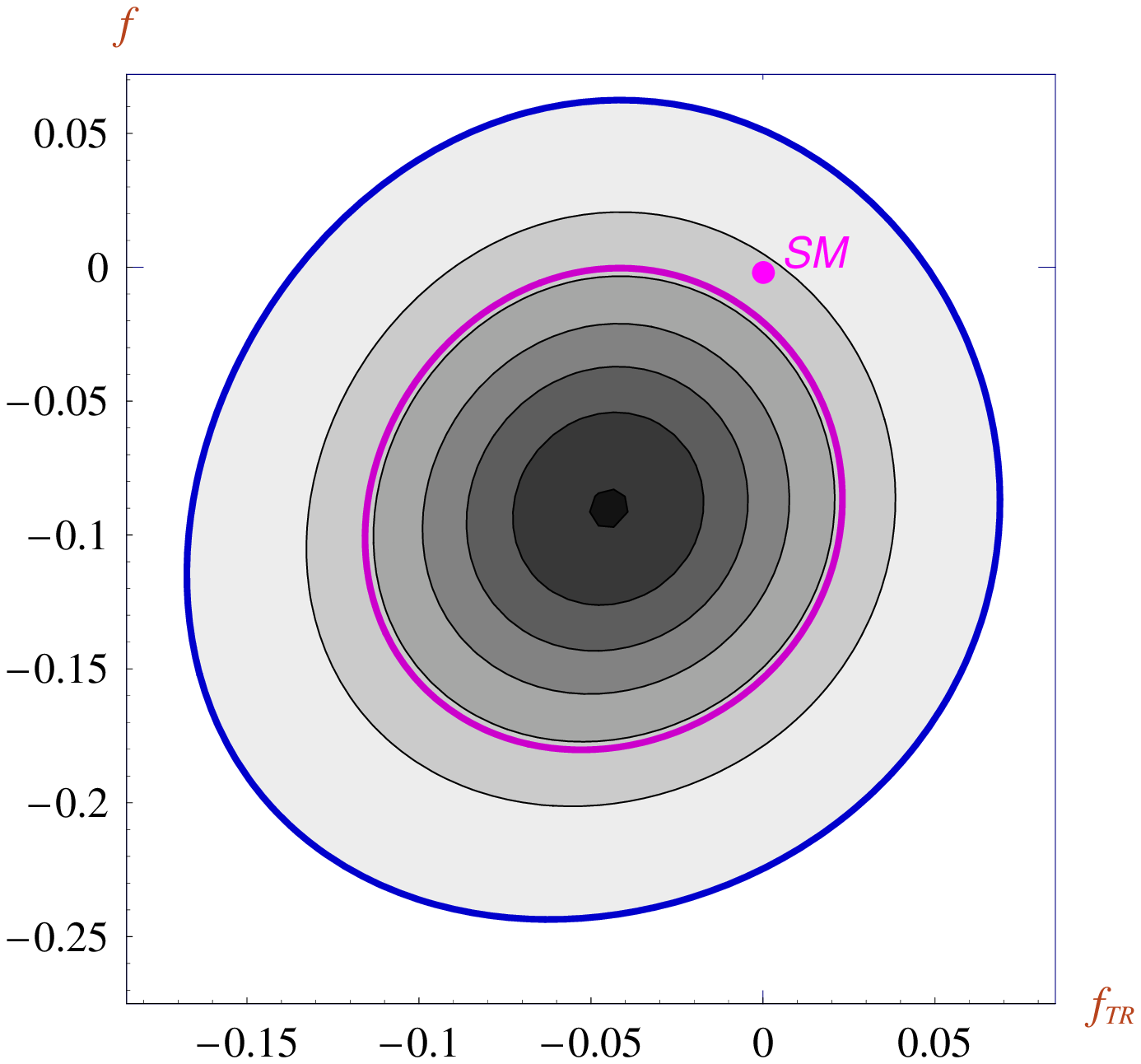} 
\end{center}
\caption{\label{2D}
The topographic maps of the 2D-distributions $\rho(f_1,f_2)$ of the anomalous couplings at the scale $\mu=2M_W$, obtained by integrating the 3D distribution over the third coupling.
(a): $f_{VL}$-$f_{TR}$; (b): $f_{VL}$-$f$; (c): $f_{TR}$-$f$.
At each plot, the blue external contour embraces the 95\% CL region; 
the violet contour embraces the 68\% CL region. The black contours 
correspond to the values $\rho(f_1,f_2)=10,20,30,\dots$.
}
\end{figure}

Integrating these distributions over one of the couplings, we obtain normalized 2D distributions 
of two other couplings $\rho(f_1,f_2)$ shown in Fig.~\ref{2D}.
In all cases, the SM values belong to the 95\% CL; however, they are outside the 68\% CL area.  

\begin{figure}[t!]
\begin{center}
\includegraphics[width=7cm]{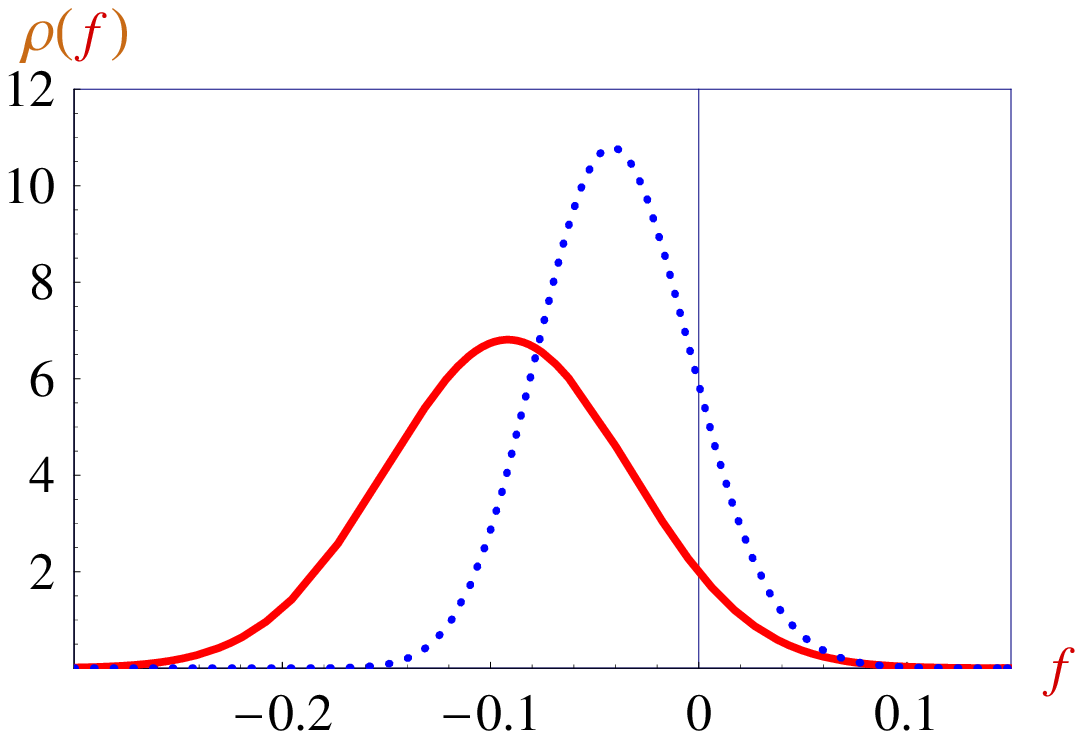} 
\includegraphics[width=7cm]{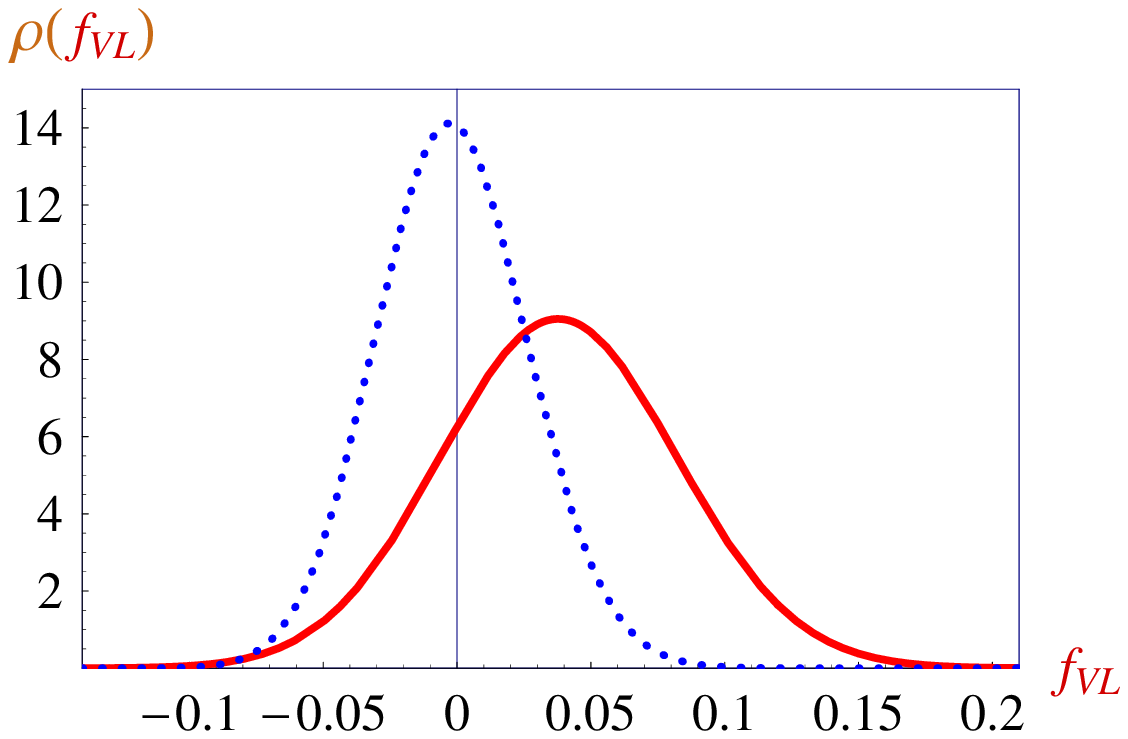}
\includegraphics[width=7cm]{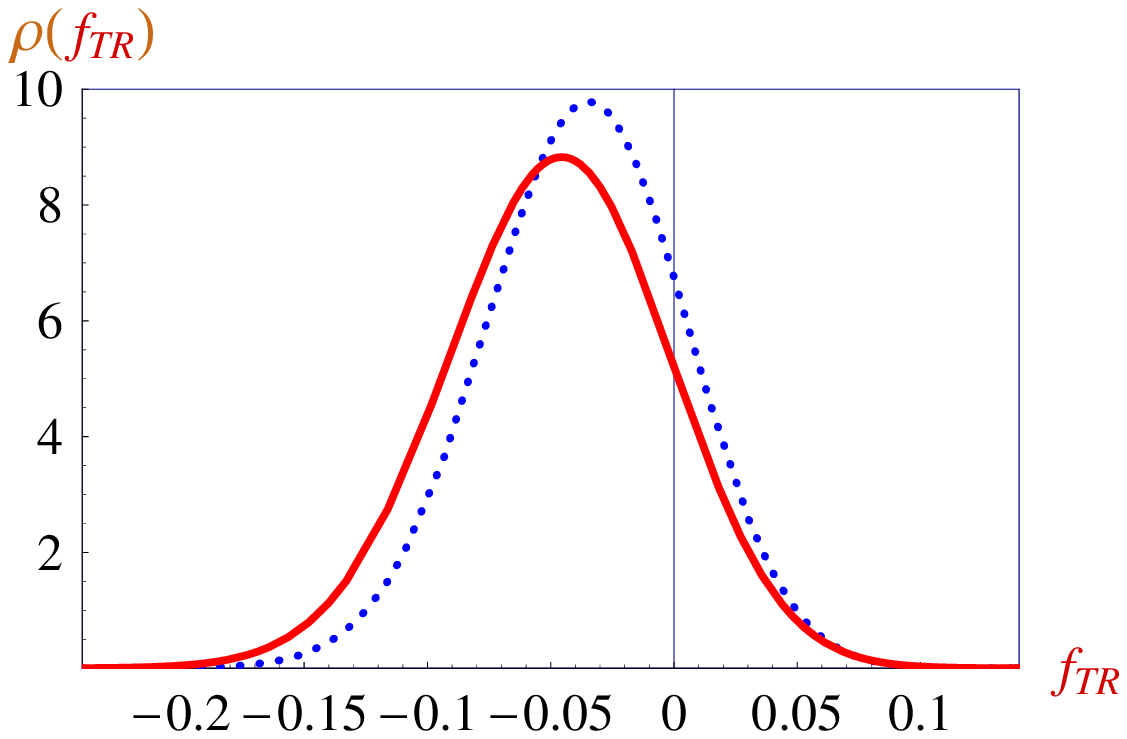}
\end{center}
\caption{\label{1D}
The 1D distributions obtained by integrating the 3D distributions (red solid lines). (a): $f$; (b): $f_{VL}$; (c): $f_{TR}$.
For comparison, the blue dashed lines show the distributions obtained by setting all other anomalous couplings to zero.
The results present the distributions of the anomalous couplings at the matching scale $\mu=2M_W$.}
\end{figure}
Finally, Fig.~\ref{1D} gives a one-dimensional distribution of the individual coupling, obtained by integrating the 3D distribution 
over other couplings.
We emphasize that these 1D distributions differ from the 1D distributions obtained by setting all other couplings to their SM values. 


\section{Discussion and conclusions}

We presented a detailed analysis of the anomalous $Wtb$ couplings based on a broad range of $B$-physics data allowing all of them to differ simultaneously from zero. Our conclusions may be summarized as follows: 

\vspace{.2cm}
\noindent 
1. Taking into account that any analysis of the $B$-physics data involves the theoretical calculation of complicated 
nonperturbative QCD effects (e.g., related to $B$-meson in the initial state, light mesons in the final state, 
charming-loops and charmonia resonances) only those modes where such effects may be controlled with good accuracy 
are appropriate for the analysis of the anomalous $Wtb$ couplings. Presently, such effects are limited to FCNC radiative 
or semileptonic $B$-decays in the region of small momentum transfers of the $l^+l^-$ pair, purely leptonic $B$-decays 
and the oscillation of neutral $B$-mesons. 
In other interesting processes/kinematic regions, where the gluon penguin operator $O_8$ or four-quark operators provide 
sizeable or dominant contributions, the nonperturbative QCD 
effects are very difficult to calculate with good accuracy; therefore such processes are not suitable for the analysis of the anomalous 
couplings from the data. Consequently, $B$-decays can provide bounds on three quantities: the anomalous couplings 
$f_{VL}$, $f_{TR}$ and one linear combination $f$ of two other couplings, $f_{VR}$ and $f_{TL}$, that appears in the Wilson 
coefficient $\delta C_7$. 

\vspace{.2cm}
\noindent 
2. Allowing simultaneous deviations of all anomalous couplings from zero and calculating the 1D distributions 
by integrating the 3D distributions leads to rather visible differences from the 1D distributions obtained by allowing 
only one anomalous coupling to take a non-SM value and keeping all other couplings zero, see Fig. \ref{1D}.

\vspace{.2cm}
\noindent 
3. Figures \ref{2D} and \ref{1D} show our results for the distributions of the anomalous couplings at the scale $\mu=2M_W$. 
In all considered 2D and 1D distributions, the SM values of the couplings belong to the region allowed at the 95\% CL.
However, the SM values lie beyond the 68\% CL region of the anomalous couplings. 
For obtaining further constraints on the anomalous couplings, in particular, for constraining the couplings $f_{VR}$ and $f_{TL}$, 
separately, combining bounds on the anomalous couplings from the $B$-physics data with the direct bounds from 
single top-quark production 
is a promising route to New Physics \cite{Wtb_Mellado,Wtb_Kozachuk,Wtb_Hiller}. 
The direct constraints on the anomalous couplings from LHC data depend on the scenarios of the anomalous couplings (1D or 2D) used in the analysis. Within 1D scenarios, indirect constraints from B-physics experiments are presently much more restrictive compared to the direct top-quark measurements \cite{Wtb_Kozachuk}.

\section{Acknowledgment}

We are grateful to 
C.~Bobeth, E.~Boos, V.~Bunichev, M.~Dubinin, L.~Dudko, P.~Mandrik, D.~Savin, D.~van Dyk for stimulating and interesting discussions. 
We thank the Organizers of the MIAPP programme ``Deciphering Strong-Interaction Phenomenology through Precision Hadron-Spectroscopy'' 
held on 7-31 October 2019 at the Excellence Cluster ``Universe'' in Garching, Germany, 
for financial support of our participation at this workshop, where a major part of this work was done. 
The work was supported by the Russian Foundation of Basic Research under grant 19-52-15022 (A.K. and D.M.) 
and by the Russian Science Foundation under grant RNF-16-12-10280 (A.K., Section 3). 



\end{document}